\documentclass[sigconf]{acmart}
\renewcommand\footnotetextcopyrightpermission[1]{}

\usepackage{adjustbox}
\usepackage{multirow}
\usepackage{pifont}
\usepackage{colortbl}
\usepackage{tabularx}
\usepackage{enumitem}
\usepackage{tcolorbox}
\usepackage[absolute,overlay]{textpos}

\AtBeginDocument{%
  }

\setcopyright{acmlicensed}
\copyrightyear{2018}
\acmYear{2018}
\acmDOI{XXXXXXX.XXXXXXX}
\acmConference[Conference acronym 'XX]{Make sure to enter the correct
  conference title from your rights confirmation email}{June 03--05,
  2018}{Woodstock, NY}
\acmISBN{978-1-4503-XXXX-X/2018/06}

\begin{document}




\settopmatter{printacmref=false}

\newcommand{\stitle}[1]{\vspace*{0.5em}\noindent{\bf #1\/}}
\newcommand{\squishlist}{
	\begin{list}{$\bullet$}
		{ \setlength{\itemsep}{1pt}
			\setlength{\parsep}{1pt}
			\setlength{\topsep}{2.5pt}
			\setlength{\partopsep}{0.5pt}
			\setlength{\leftmargin}{1em}
			\setlength{\labelwidth}{1em}
			\setlength{\labelsep}{0.6em}
		}
	}
	\newcommand{\squishend}{
	\end{list}
}

\title{UniVA: Unified Value Alignment for Generative Recommendation in Online Advertising at Tencent}

\author{{\Large Xinxun Zhang $^{1*}$, Yuling Xiong $^{2*}$, Yangru Huang $^{2*}$, Jiale Zhou $^{3*}$, Zhengkai Guo $^2$, Zhennan Pang $^2$, Junbang Huo $^2$, Jingwen Wang $^2$, Xuyang Sun $^2$, Enming Zhang $^2$, Jiaguang Jin $^2$, Changping Wang $^2$, Yi Li $^2$, Jun Zhang $^{2\dagger}$, Xiao Yan $^1$, Jiawei Jiang $^{1\dagger}$, Jie Jiang $^{2\dagger}$,}}

\affiliation{
  \institution{$^1$ Wuhan University, China; $^2$ Tencent Inc., China; $^3$ Peking University, China}
  \country{}
}
\email{{xxzhangstu, jiawei.jiang, yanxiaosunny}@whu.edu.cn}
\email{jlzhou25@stu.pku.edu.cn, {whitnyxiong, yarayrhuang, neoxzhang, zeus}@tencent.com}

\renewcommand{\shortauthors}{Zhang et al.}



\begin{abstract}
Generative Recommendation (GR) reformulates recommendation as next-token generation over item Semantic IDs (SIDs) and has shown promise in industrial applications. However, extending GR to advertising is non-trivial, as advertising recommendation must jointly account for user relevance and commercial value. This creates a mismatch: high generation likelihood does not necessarily imply high advertising utility. As a result, valuable ads may be poorly distinguished in the SID space, pruned during autoregressive decoding, or missed when request-invalid branches consume limited beam capacity during online serving. To address this problem, we propose UniVA, a \textbf{Uni}fied \textbf{V}alue \textbf{A}lignment framework for generative advertising recommendation. {UniVA aligns commercial value across the entire pipeline of SID construction, autoregressive decoding, and online serving.} Commercial SID Tokenization injects business attributes and bid information into SID construction. A Generation-as-Ranking SID Decoder then fuses generation scores with token-level value estimates during autoregressive decoding. {Finally, Value-Aware Constrained Serving restricts the fused decoding process to request-valid SID paths through a personalized trie.} Experiments on the Tencent WeChat Channels advertising platform show that UniVA achieves a 37.04\% relative improvement in offline Hit Rate@100 over the baseline and lifts gross merchandise value (GMV) by 1.5\% in online A/B tests.
\end{abstract}



\maketitle

\def\thefootnote{*}\footnotetext{These authors contributed equally to this work}
\def\thefootnote{$\dagger$}\footnotetext{Corresponding author}

\section{Introduction}

Generative Recommendation (GR) represents each item as a discrete Semantic ID (SID) sequence and reformulates recommendation as autoregressive SID generation conditioned on user history \cite{rajput2023recommender,hou2023learning,zhang2025gpr}. By replacing full-corpus scoring with SID generation, GR unifies user behavior modeling, candidate generation, and item retrieval within one framework and has shown strong potential in search, e-commerce, and content recommendation \cite{li2025survey,li2024survey,wang2025nezha,zhou2025onerec,hao2025oxygenrec}. Since candidate selection is determined by generation probability, GR implicitly assumes that the generation likelihood of an SID path is aligned with recommendation utility.

Advertising recommendation, however, breaks this alignment. It is inherently multi-objective: beyond user relevance, the system must also account for the commercial return of an advertisement under the current request. Such value depends on advertiser bids, return-on-investment (ROI) targets, expected cost per mille (eCPM), and other business-side factors \cite{jiang2022adaptive,pei2019value,zheng2025ega,zhao2025llm}. Generation likelihood alone therefore cannot fully characterize advertising utility, making commercial value modeling essential for generative advertising recommendation.

Existing generative advertising recommendation methods have recognized the importance of commercial value and incorporated auction signals, business-side supervision, or value-oriented objectives into GR frameworks \cite{zhang2025gpr,zheng2025ega,xue2026GR4AD,xiong2026llatte}. These signals are typically introduced through auxiliary objectives, sample reweighting, or value-aware adjustments at specific stages of training or inference. While such designs improve value sensitivity locally, the overall generative process remains largely likelihood-centered, without consistent value alignment across the entire GR pipeline. Consequently, valuable ad candidates may still be under-represented, pruned, or omitted during generation.

\begin{figure}[t]
\centering
\includegraphics[width=0.95\linewidth]{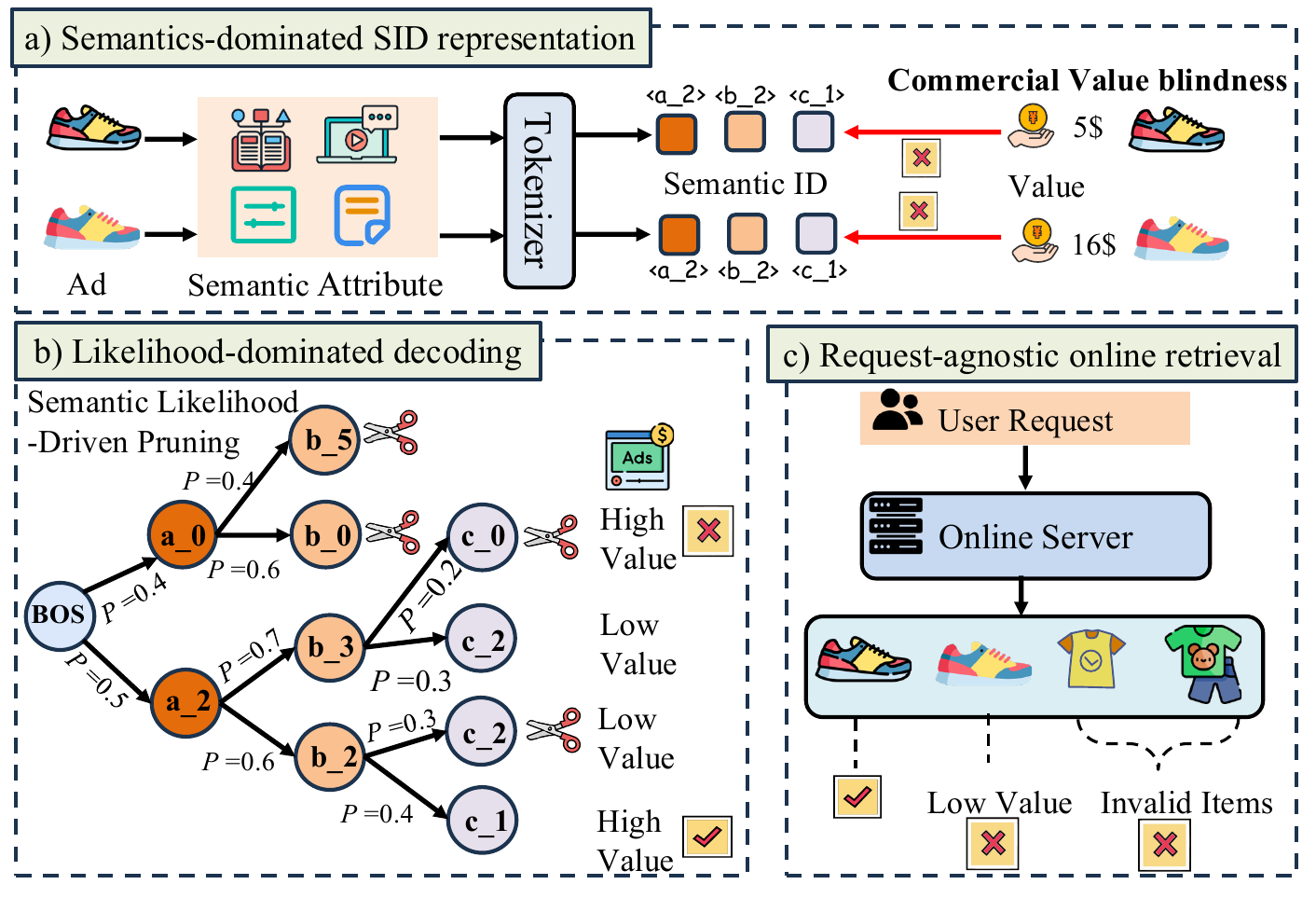} 
\caption{Existing methods can lose commercial value at three stages: (a) semantic SID tokenization may conflate ads with different commercial attributes, (b) likelihood-dominated decoding can prune low-probability but high-value SID trajectories, and (c) request-agnostic retrieval wastes beam search capacity on low-value or request-ineligible items.}
\label{fig_motivation}    
\end{figure}

As illustrated in Figure~\ref{fig_motivation}, commercial value can be lost at three key stages of generative advertising recommendation. \textbf{\ding{182} Semantics-centric SID representation.} Existing SID tokenization methods primarily organize ads according to semantic similarity. Consequently, ads with similar content but substantially different bids, ROI targets, or conversion values may be mapped to nearby SID paths, and the resulting token space has limited discriminability with respect to commercial value. \textbf{\ding{183} Likelihood-dominated autoregressive decoding.} An SID path is generated token by token. If a high-value candidate has a low generation likelihood on its early prefix, it may be prematurely pruned during beam expansion. Once pruned, the candidate is no longer available to subsequent ranking or value estimation. \textbf{\ding{184} Request-agnostic online retrieval.} The set of eligible ads varies across requests, whereas conventional SID decoding searches over a global path space independent of the current request. Consequently, request-invalid branches consume computation and limited beam capacity, and eligible high-value candidates may be excluded from the final result.

\stitle{Our Solution.} To address these challenges, we propose \textbf{UniVA}, a \textbf{Uni}fied \textbf{V}alue \textbf{A}lignment framework for GR in industrial advertising. {UniVA integrates commercial value modeling throughout the entire GR pipeline, spanning SID construction, autoregressive decoding, and online serving.}

UniVA first builds a value-discriminative token space through \textbf{Commercial SID (CSID) Tokenization}, forming a semantic--commercial hierarchy. The upper SID levels preserve coarse semantic organization, while the final token combines structured business attributes with discretized bid information. This structure distinguishes ads with similar semantics but different commercial values without disrupting the upper-level hierarchy. Based on the token space, the \textbf{Generation-as-Ranking SID Decoder} uses a shared decoder trunk with a generation head and an action-value head. The generation head predicts next-token likelihoods over the SID vocabulary, while the action-value head estimates the commercial return of candidate token expansions. Supervised and eCPM-aware reinforcement learning jointly train the two heads, whose outputs are fused during online decoding to let commercial value guide prefix decisions before promising trajectories are pruned. Finally, \textbf{Value-Aware Constrained Serving} executes the fused policy over a request-specific valid-path trie. The trie masks expansions that violate inventory or targeting constraints, while the fused logits rank the remaining actions. UniVA thereby performs generation, ranking, and validity filtering in one decoding process without a separate online value model or post-generation reranker.

We conduct systematic evaluations on two public benchmarks and the Tencent WeChat Channels advertising platform. On the public benchmarks, UniVA consistently outperforms all baselines, with relative improvements of up to 8.4\% in HR and 7.8\% in NDCG. On the industrial advertising benchmark, UniVA achieves a 37.04\% relative improvement in HR@100 over the strongest GR baseline. It also improves ValueHR@100 and wNDCG@100, while reducing bid dispersion within each SID path by about one order of magnitude. Finally, online A/B tests on 20\% of production traffic yield a 1.50\% GMV lift, validating the effectiveness of UniVA in industrial deployment. The main contributions of this work are as follows:

\begin{itemize}[leftmargin=*]
\item We characterize value inconsistency in generative advertising recommendation as a three-stage problem spanning SID representation, autoregressive decoding, and online serving.
\item We introduce a semantic--commercial SID structure that separates semantic organization from commercial differentiation, enabling semantically similar ads with different commercial values to follow distinct SID paths.
\item We formulate SID decoding as joint generation and ranking, allowing commercial value to guide prefix-level decisions before high-value candidates are pruned.
\item {We design Value-Aware Constrained Serving to restrict fused decoding to request-valid SID paths, avoiding invalid expansions without a separate online reranking stage.}
\end{itemize}

\section{Preliminaries}
\label{sec:preliminaries}
{
\stitle{SID-based Generative Recommendation.}
Given an advertising request $x=(u,c,\mathbf{x}_{1:T})$, generative recommendation directly retrieves the next item by generating its discrete identifier. Here, $u$ denotes the user, $c$ the request context, and $\mathbf{x}_{1:T}$ the historical interaction sequence. Each advertisement $i$ is mapped to a length-$L$ Semantic ID (SID) sequence $s_i=\Phi(i)=(s_i^1,\ldots,s_i^L)$. Residual-quantization (RQ)-based tokenizers organize these tokens into a coarse-to-fine semantic hierarchy \cite{rajput2023recommender,zhang2025gpr}. For a generated trajectory $y=(a_1,\ldots,a_L)$ with prefix $s_{<l}=(a_1,\ldots,a_{l-1})$, the autoregressive policy factorizes as $\pi_\theta(y\mid x)=\prod_{l=1}^{L}\pi_\theta(a_l\mid x,s_{<l})$.

\stitle{Advertising Attributes and Value Signals.}
Each advertisement contains semantic attributes $x_i^{s}=(x_i^{\text{text}},x_i^{\text{img}},x_i^{\text{video}})$ and commercial attributes $x_i^{c}=(x_i^{o},x_i^{r},x_i^{\mathrm{ind}},x_i^{b})$, corresponding to optimization goal, return-on-investment (ROI) target, industry, and bid. UniVA distinguishes three value-related signals. The commercial attributes describe an ad and support SID construction, but are not reward labels. A complete feasible trajectory receives a terminal reward $R_{\mathrm{eCPM}}(x,y)$ from the production-style ranking stack. UniVA further learns a request-conditioned token-level value estimate $q_\phi(s_{<l},a;x)$ for the downstream commercial return of each candidate action.

\stitle{Problem Definition.}
{At serving time, inventory, targeting, and creative constraints restrict generation to a request-specific feasible set $\mathcal{Y}(x)$. The most likely trajectory under the generative policy may not yield the highest commercial return. UniVA therefore learns two complementary functions. The generation policy $\pi_\theta$ retains supervised next-SID behavior while being optimized toward higher expected terminal eCPM over feasible trajectories. In parallel, the action-value function $q_\phi(s_{<l},a;x)$ estimates the downstream return of each candidate token expansion. During online inference, UniVA fuses the generation scores from $\pi_\theta$ with $q_\phi$ and restricts decoding to $\mathcal{Y}(x)$. This formulation separates policy optimization, action-value regression, and serving-time fusion while aligning all three with the same terminal commercial objective.}
}
 
\section{Methodology}
\label{sec:method}

\begin{figure*}[htbp]
\centering
\includegraphics[width=0.95\linewidth]{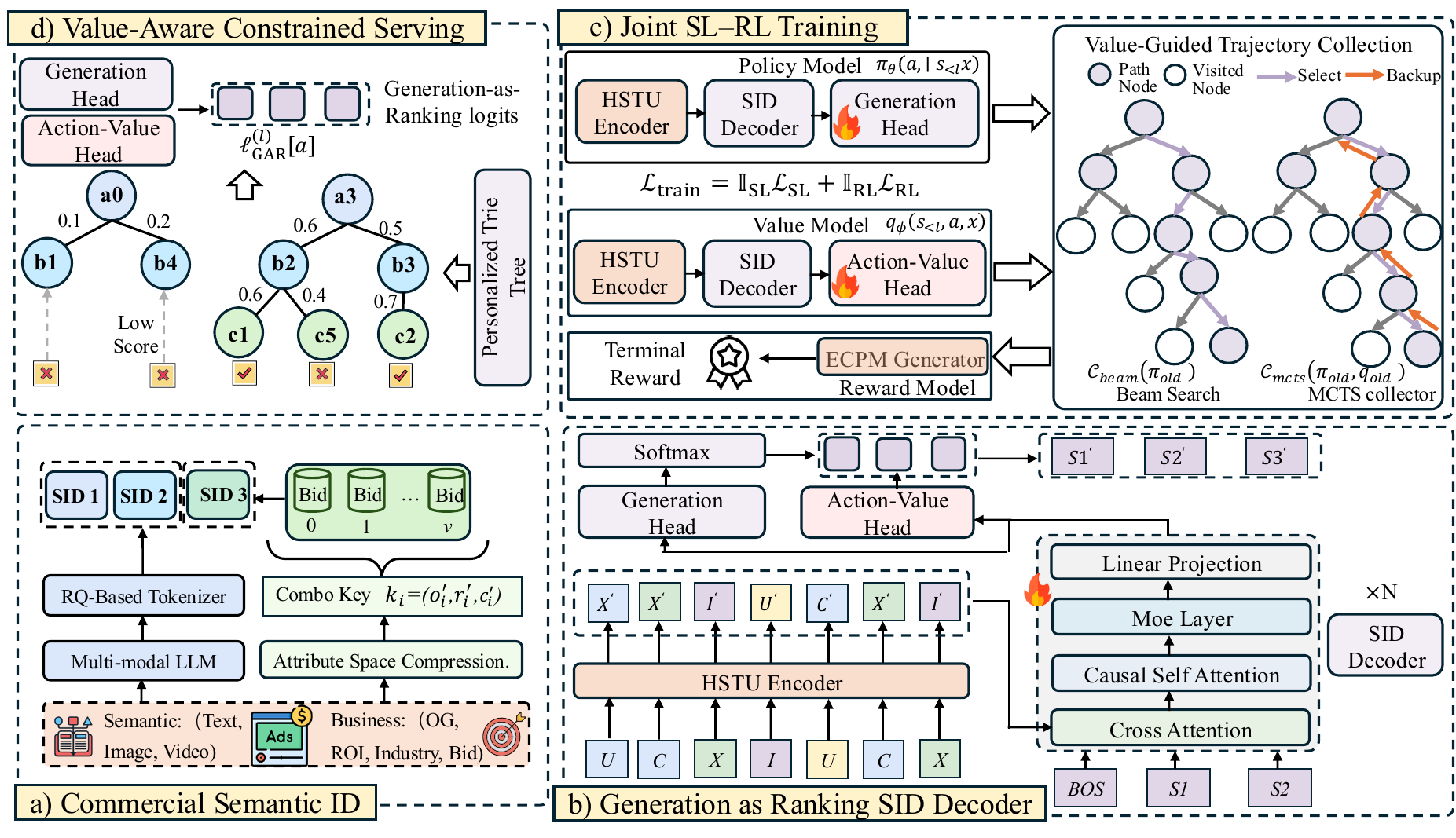}
\caption{Overview of UniVA. CSID Tokenization builds a value-discriminative action space. A shared two-head decoder is trained through supervised learning and eCPM-aware RL with trajectories collected by beam search and Monte Carlo Tree Search (MCTS). During online serving, its generation logits and learned action values are fused, while a request-specific trie constrains beam search to valid SID paths.}
\label{fig_model}
\end{figure*}

Figure~\ref{fig_model} presents an overview of UniVA, which aligns commercial value across offline training and online inference. During offline training, CSID Tokenization preserves the semantic hierarchy of SIDs while distinguishing ads with different commercial values. The SID decoder learns two complementary signals over a shared backbone. Its generation head defines the autoregressive policy and is optimized by supervised learning and eCPM-aware reinforcement learning, while its action-value head learns the expected commercial return of candidate token expansions and guides offline trajectory collection. {During online inference, the Generation-as-Ranking decoder fuses generation logits with action-value estimates for token selection. Value-Aware Constrained Serving then applies Value-Guided Personalized Beam Search to restrict these fused logits to request-valid SID paths.}

\subsection{CSID Tokenization}
\label{sec:commercial_sid}

Existing SID construction methods primarily encode semantic characteristics, which leaves the token space insufficiently discriminative for commercial value. To address this limitation, UniVA constructs a semantic--commercial SID hierarchy:

\begin{equation}
(s_i^{1}, \dots, s_i^{L-1}) = \Phi_{\mathrm{sem}}(x_i^{s}), \qquad s_i^{L} = \Phi_{\mathrm{com}}(x_i^{c}),
\end{equation}

where $\Phi_{\mathrm{sem}}$ reuses the RQ-KMeans+ semantic tokenizer \cite{zhang2025gpr} to preserve the semantic organization of the upper SID levels, and $\Phi_{\mathrm{com}}$ maps structured commercial attributes into a discrete value-aware token at the final level. In this way, UniVA preserves the coarse-to-fine semantic hierarchy of GR while explicitly injecting commercial discriminability into the SID path. The construction of $\Phi_{\mathrm{com}}$ consists of two steps: attribute space compression, followed by composition-key construction with equal-frequency bid binning.

\stitle{Attribute Space Compression.}
The Cartesian product of raw commercial attributes is high-cardinality and sparse, making it difficult to construct statistically reliable commercial tokens directly. UniVA therefore compresses the optimization goal, ROI target, and industry attribute independently before composing them. Formally, we define
\begin{equation}
	x_i^{o'} = \phi_{\mathcal{O}}(x_i^{o}), \qquad x_i^{r'} = \phi_{\mathcal{R}}(x_i^{r}), \qquad x_i^{\mathrm{ind}'} = \phi_{\mathcal{I}}(x_i^{\mathrm{ind}}),
\end{equation}

where $\phi_{\mathcal{O}}$, $\phi_{\mathcal{R}}$, and $\phi_{\mathcal{I}}$ denote the compression operators for optimization goal, ROI, and industry, respectively. Each operator preserves frequent attribute values and maps long-tail values into a compact set of groups. For optimization goals, tail values are clustered according to the similarity of their bid distributions. For ROI and industry, tail values are merged into fallback categories. This compression reduces sparsity in the subsequent composition space and keeps the commercial token vocabulary manageable. Dataset-specific thresholds and resulting category counts are reported in Appendix~\ref{app:implementation_details}.

\stitle{Value-Aware Discretization.}
After attribute compression, UniVA constructs a composition key for each advertisement from the compressed business attributes:

\begin{equation}
k_i = (x_i^{o'}, x_i^{r'}, x_i^{\mathrm{ind}'})
\in \mathcal{K}
\subseteq \mathcal{O}' \times \mathcal{R}' \times \mathcal{I}'.
\end{equation}

{Each key represents a local commercial context in which ads share similar business conditions. Before bid-bin allocation, keys with fewer than $m_{\min}$ observations are merged into a reserved fallback key $k_{\mathrm{fb}}$, where $m_{\min}$ denotes the minimum sample support for retaining a composition key. The fallback key is then treated as an ordinary key during vocabulary allocation.} For each $k\in\mathcal{K}$, its bid set $\mathcal{B}_k$ and sample proportion $w_k$ are defined as

\begin{equation}
\mathcal{B}_{k} = \{x_i^{b}\mid k_i=k\}, \qquad
w_k = \frac{|\mathcal{B}_k|}{\sum_{k'\in\mathcal{K}}|\mathcal{B}_{k'}|}
\end{equation}

UniVA assigns $n_k$ bid bins to each key under a commercial-token vocabulary budget $V$. {Here $b_{\min}$ and $b_{\max}$ denote the minimum and maximum numbers of bid bins assigned to a retained key, respectively; they are distinct from the minimum key-support threshold $m_{\min}$. We choose positive integer bin counts satisfying $\sum_{k\in\mathcal K}n_k\leq V$ and $b_{\min}\leq n_k\leq\min(b_{\max},|\mathcal B_k|)$ for every $k$.} Because bids are partitioned into equal-frequency bins within each key, every key-bin token has empirical probability approximately $w_k/n_k$. {We therefore maximize the resulting commercial-token entropy $H_{\mathrm{tok}}$:}

\begin{equation}
{
\{n_k^\star\}_{k\in\mathcal K}
=\arg\max_{\{n_k\}}
\left[-\sum_{k\in\mathcal K}w_k\log\frac{w_k}{n_k}\right].
}
\end{equation}

This objective allocates more bid resolution to dense commercial contexts while preventing sparse keys from creating poorly supported tokens. After $\{n_k\}$ is fixed, UniVA partitions each $\mathcal{B}_k$ by empirical quantiles. Let $b_k(x_i^b)\in\{1,\ldots,n_k\}$ denote the resulting bin index. A global indexing function $\psi$ assigns one commercial token to each key-bin pair:

\begin{equation}
s_i^{L}
= \Phi_{\mathrm{com}}(x_i^c)
= \psi\big(k_i,b_{k_i}(x_i^b)\big).
\end{equation}

For a composition key not observed during construction, UniVA maps it to $k_{\mathrm{fb}}$, whose bins are estimated from the pooled fallback bid distribution. The resulting commercial token separates ads by both business context and relative bid level.

\stitle{SID Buckets and Ad Resolution.}
The complete CSID mapping remains many-to-one. A path $y$ therefore indexes an advertisement bucket
\begin{equation}
\mathcal{A}(y)=\{i\mid\Phi(i)=y\}.
\end{equation}
For a request $x$ with eligible inventory $\mathcal{I}(x)$, UniVA retains the eligible subset and resolves the path by bid priority:
\begin{equation}
\mathcal{A}_x(y)=\mathcal{A}(y)\cap\mathcal{I}(x), \qquad
\rho_x(y)=\arg\max_{i\in\mathcal{A}_x(y)}x_i^b.
\label{eq:leaf_resolution}
\end{equation}
This deterministic rule is used in both offline reward simulation and online serving.

\subsection{Generation-as-Ranking SID Decoder}
\label{sec:gar_decoder}
{
CSID encodes value-relevant distinctions in the discrete token space, but the preferred token still depends on the current user and request. UniVA therefore introduces a Generation-as-Ranking SID Decoder that scores each candidate token by generation relevance and request-conditioned commercial value. Following GPR \cite{zhang2025gpr}, we adopt its unified input schema and HSTU encoder backbone \cite{zhai2024actions}. The input sequence contains User Token ($U$), Organic Token ($O$), Environment Token ($E$), and Item Token ($I$), which encode user attributes and preferences, organic-content behaviors, request context, and historical ad interactions, respectively. The encoder produces contextualized states $h = \mathrm{Enc}(U,O,E,I)$.

\stitle{Scalable Decoder Backbone.}
Conditioned on $h$, the decoder autoregressively generates the target SID. Let $z_l^{(d)}$ denote the representation at SID position $l$ entering decoder block $d$. At step $l$, each block first performs fully visible cross-attention over the encoder states for the generated prefix and then applies causal self-attention:

\begin{equation}
\tilde{z}_{1:l}^{(d)} =
\mathrm{CrossAttn}(Q=z_{1:l}^{(d)},K=h,V=h),
\end{equation}
\begin{equation}
\hat{z}_l^{(d)} =
\mathrm{SelfAttn}\!\left(
Q=\tilde{z}_l^{(d)},
K=\tilde{z}_{1:l}^{(d)},
V=\tilde{z}_{1:l}^{(d)}
\right).
\end{equation}

Cross-attention injects request-aware user context, while self-attention captures dependencies among previously generated SID tokens. To provide sufficient policy capacity under heterogeneous advertising distributions, UniVA combines sparse Mixture-of-Experts (MoE) \cite{dai2024deepseekmoe} with recursive parameter sharing inspired by Mixture-of-Recursions (MoR) \cite{bae2025mixture}. For MoE, UniVA activates only the top-$K_{\mathrm{exp}}$ routed experts for each token:

\begin{equation}
g(\hat{z}_l^{(d)}) = \mathrm{Softmax}(W_r\hat{z}_l^{(d)}),
\qquad
\mathcal{M}_l^{(d)}
=\mathrm{TopK}\!\left(g(\hat{z}_l^{(d)}),K_{\mathrm{exp}}\right),
\end{equation}
\begin{equation}
\bar{g}_m(\hat{z}_l^{(d)})
=\frac{g_m(\hat{z}_l^{(d)})}
{\sum_{j\in\mathcal{M}_l^{(d)}}g_j(\hat{z}_l^{(d)})},
\qquad m\in\mathcal{M}_l^{(d)},
\end{equation}
\begin{equation}
z_l^{(d+1)}
=E_0(\hat{z}_l^{(d)})
+\sum_{m\in\mathcal{M}_l^{(d)}}
\bar{g}_m(\hat{z}_l^{(d)})E_m(\hat{z}_l^{(d)}).
\end{equation}

Here $E_0$ is an always-activated shared expert, whereas the routed experts specialize in context-dependent patterns. Dynamic load balancing adjusts routing bias using historical expert-load statistics, reducing router collapse and expert-capacity waste.

For depth scaling, UniVA organizes these decoder blocks into an MoR-style recurrent computation with a shared middle transformation and a fixed recursion budget $R$. Let $e_l$ be the input representation at SID position $l$. The final decoder state is

\begin{equation}
u_l^{(0)}=\ell_{\mathrm{in}}(e_l,h), \qquad
u_l^{(r)}=\ell_{\mathrm{mid}}(u_l^{(r-1)},h), \qquad
z^{(l)}=\ell_{\mathrm{out}}(u_l^{(R)}).
\end{equation}

Together, sparse MoE expands conditional capacity across heterogeneous advertising regimes, while recursive parameter sharing increases effective depth without allocating an independent parameter set at every recursion. The resulting backbone provides a scalable context-conditioned representation for the subsequent Generation-as-Ranking policy.

\stitle{Generation and Action-Value Heads.}
On top of the shared decoder trunk, UniVA introduces a generation head and an action-value head. Let $z^{(l)}$ denote the decoder hidden state and $\mathcal{S}_l$ the SID vocabulary at level $l$. The generation head defines the autoregressive policy, whereas the action-value head estimates the downstream commercial return of each candidate token:

\begin{equation}
\begin{array}{l}
	o_{\mathrm{gen}}^{(l)} = f_{\mathrm{gen}}(z^{(l)}), \qquad
	o_{\mathrm{value}}^{(l)} = f_{\mathrm{value}}(z^{(l)}), \\
	\pi_{\theta}(\cdot \mid s_{<l}, h)
	= \mathrm{Softmax}\left(o_{\mathrm{gen}}^{(l)}\right), \\
	q_\phi(s_{<l},a;x)
	= o_{\mathrm{value}}^{(l)}[a], \qquad a\in\mathcal{S}_l.
\end{array}
\label{eq:generation_value_heads}
\end{equation}

{Here $f_{\mathrm{gen}}(\cdot)$ and $f_{\mathrm{value}}(\cdot)$ denote the two output heads, and $o_{\mathrm{gen}}^{(l)}$ and $o_{\mathrm{value}}^{(l)}$ are their vocabulary-level outputs at SID level $l$. The generation policy $\pi_\theta$ models user relevance and autoregressive SID compatibility. The action-value function $q_\phi$ predicts the request-conditioned return of selecting token $a$ at the current prefix. The two heads share the decoder representation but have distinct roles during training. The Proximal Policy Optimization (PPO) term \cite{schulman2017proximal} transfers terminal eCPM feedback into $\pi_\theta$, while return regression trains $q_\phi$. The behavior-policy expectation of these action values provides the PPO state baseline, and the individual action values serve as the search heuristic.}

\stitle{Generation-as-Ranking Fusion.}
During inference, UniVA casts each next-token decision as ranking the candidate SID expansions by both generation relevance and predicted commercial return. At SID level $l$, the two outputs are fused as
\begin{equation}
\ell_{\mathrm{GAR}}^{(l)}[a]
=o_{\mathrm{gen}}^{(l)}[a]
+\alpha_l q_\phi(s_{<l},a;x),
\qquad a\in\mathcal{S}_l,
\label{eq:generation_value_fusion}
\end{equation}
where $\alpha_l\geq 0$ is a level-specific calibration coefficient selected on held-out validation data and fixed during online serving. Setting $\alpha_l=0$ recovers the RL-trained generation policy, whereas $\alpha_l>0$ enables Generation-as-Ranking by allowing predicted commercial return to directly affect token selection. {The same action-value head trained by return regression within the joint RL objective is therefore reused during online decoding, without introducing a separate ranking model.}

}

\subsection{Joint SL--RL Training}
\label{sec:ecpm_rl}
{
During offline training, UniVA fixes the CSID construction and trains the two-head decoder with complementary supervised and reinforcement learning signals. Supervised learning establishes valid and user-relevant SID generation, while eCPM-aware reinforcement learning refines the generation policy and action-value estimates using terminal commercial feedback.

\stitle{Supervised Generation Learning.}
For a supervised sample $(x,s^\star)\in\mathcal{D}_{\mathrm{SL}}$, the next-SID objective is
\begin{equation}
\mathcal{L}_{\mathrm{SL}} = - \sum_{(x,s^\star)\in\mathcal{D}_{\mathrm{SL}}}\sum_{l=1}^{L}\log \pi_\theta(s_l^\star\mid x,s_{<l}^\star).
\end{equation}
This objective updates the shared decoder trunk and generation head, anchoring the policy to stable SID generation before and during reinforcement learning.
 
\stitle{Value-Guided Trajectory Collection.}
Likelihood-driven beam search covers high-probability SID trajectories but may miss low-probability prefixes with high terminal return. UniVA therefore combines beam rollouts with a Monte Carlo Tree Search (MCTS) collector inspired by value-guided structured sampling \cite{jiang2026spend}:
\begin{equation}
\begin{aligned}
\mathcal{C}(x) =\;&
\mathcal{C}_{\mathrm{beam}}\big(\pi_{\mathrm{old}}\big)
\cup
\mathcal{C}_{\mathrm{mcts}}\big(\pi_{\mathrm{old}},q_{\mathrm{old}}\big) \\
=\;&\{y^{(1)},\ldots,y^{(K_{\mathrm{traj}})}\}.
\end{aligned}
\end{equation}
Here $\mathcal{C}_{\mathrm{beam}}$ and $\mathcal{C}_{\mathrm{mcts}}$ denote the trajectory sets returned by beam search and MCTS, respectively, and $K_{\mathrm{traj}}$ is the total number of collected trajectories. The symbols $\pi_{\mathrm{old}}$ and $q_{\mathrm{old}}$ denote frozen snapshots of the generation policy and action-value function at the beginning of a collection round. Beam search provides stable coverage of trajectories preferred by the behavior policy, while MCTS uses $q_{\mathrm{old}}$ to allocate additional search to value-promising prefixes.

Each collected trajectory is resolved to a concrete advertisement and evaluated by an offline simulator built from recent production snapshots following GPR \cite{zhang2025gpr}. The resulting eCPM is normalized within the request and used as terminal feedback for policy and action-value learning. The simulator and MCTS are used only offline. Reward construction, the MCTS search rule, and value backup are detailed in Appendix~\ref{app:optimization_details}.

\stitle{eCPM-aware Policy Optimization.}
{Group Relative Policy Optimization (GRPO) \cite{shao2024deepseekmath} is a common critic-free choice for reward alignment in generative models. UniVA instead adopts PPO because its explicit action-value learning provides prefix-level estimates that can be reused by both offline MCTS and online Generation-as-Ranking; Appendix~\ref{app:optimization_details} further discusses this choice.} UniVA optimizes the generation policy using PPO and trains the same action-value head against token-level return targets derived from the normalized terminal reward. For each prefix, the state baseline is obtained by taking the behavior policy expectation of $q_{\mathrm{old}}$ over valid next-token actions, avoiding a separate value network. Let $\widehat{A}_l$ and $\widehat{G}_l$ denote the resulting advantage and fixed return target. The RL objective is
\begin{equation}
\mathcal{L}_{\mathrm{RL}}
=\mathcal{L}_{\mathrm{PPO}}
 +\lambda_v\,
\mathbb{E}\left[
\left(q_\phi(s_{<l},a_l;x)-\widehat{G}_l\right)^2
\right].
\end{equation}
The PPO term transfers eCPM feedback into the generation policy, while the return-regression term trains the action values reused by both MCTS and Generation-as-Ranking. Thus, the action-value head is not a training-only critic: the same vocabulary-level estimates support offline exploration and are retained for online token ranking. The advantage construction and clipped PPO objective are given in Appendix~\ref{app:optimization_details}.

\stitle{Alternating Updates.}
UniVA alternates supervised and reinforcement learning batches rather than treating them as independent training stages. On RL batches, the policy loss updates the shared decoder representation and generation head, while the value loss updates the shared representation and action-value head. The batch-level training objective is
\begin{equation}
\mathcal{L}_{\mathrm{train}} = \mathbb{I}_{\mathrm{SL}}\mathcal{L}_{\mathrm{SL}} + \mathbb{I}_{\mathrm{RL}}\mathcal{L}_{\mathrm{RL}},
\end{equation}
where $\mathbb{I}_{\mathrm{SL}}$ and $\mathbb{I}_{\mathrm{RL}}$ indicate supervised and RL batches, respectively. Alternating the two objectives prevents terminal-reward optimization from drifting away from valid SID generation, while allowing commercial feedback to refine token-level decisions. The updated action-value head improves the next round of value-guided trajectory collection. PPO is defined over $\pi_\theta$, and {the same jointly trained action-value head is reused for online Generation-as-Ranking. Section~\ref{sec:value_aware_serving} describes how the resulting fused logits are used under online serving constraints.}
}

\subsection{Value-Aware Constrained Serving}
\label{sec:value_aware_serving}
{
{Online serving applies the Generation-as-Ranking logits only to request-valid SID trajectories.} Inventory, targeting, availability, and creative rules determine which SID trajectories can resolve to deliverable advertisements. UniVA therefore combines the trained two-head decoder with a personalized valid-path trie. {The trie removes infeasible actions, while the generation--value logits rank the remaining actions.}

\begin{table*}[t]
\centering
\caption{Overall performance comparison on the public Amazon Industrial and Office datasets in terms of HR@K and NDCG@K. Best results are \textbf{bolded} and second-best results are {underlined}.}
\label{tab:main_result}

\setlength{\tabcolsep}{2.8pt}
\begin{tabular}{llccccccccccc}
\toprule
\textbf{Dataset} & \textbf{Metric} & \textbf{GRU4Rec} & \textbf{Caser} & \textbf{SASRec} & \textbf{HSTU} & \textbf{TIGER} & \textbf{LCRec} & \textbf{BIGRec} & \textbf{$D^3$} & \textbf{S-DPO} & \textbf{MiniOneRec} & \textbf{UniVA} \\
\midrule
\multirow{6}{*}{\textbf{Industrial}}
& HR@3 & 0.0638 & 0.0618 & 0.0790 & 0.0927 & 0.0852 & 0.0915 & 0.0931 & 0.1024 & 0.1032 & \underline{0.1143} & \textbf{0.1160} \\
& NDCG@3 & 0.0542 & 0.0514 & 0.0700 & 0.0885 & 0.0742 & 0.0805 & 0.0841 & 0.0991 & 0.0906 & \underline{0.1011} & \textbf{0.1031} \\
& HR@5 & 0.0774 & 0.0717 & 0.0909 & 0.1037 & 0.1010 & 0.1057 & 0.1092 & 0.1213 & 0.1238 & \underline{0.1321} & \textbf{0.1359} \\
& NDCG@5 & 0.0598 & 0.0555 & 0.0748 & 0.0918 & 0.0807 & 0.0862 & 0.0907 & 0.0989 & 0.0991 & \underline{0.1084} & \textbf{0.1113} \\
& HR@10 & 0.0999 & 0.0942 & 0.1088 & 0.1163 & 0.1321 & 0.1332 & 0.1370 & 0.1500 & 0.1524 & \underline{0.1586} & \textbf{0.1632} \\
& NDCG@10 & 0.0669 & 0.0628 & 0.0806 & 0.0958 & 0.0908 & 0.0952 & 0.0997 & 0.1082 & 0.1082 & \underline{0.1167} & \textbf{0.1202} \\
\midrule
\multirow{6}{*}{\textbf{Office}}
& HR@3 & 0.0629 & 0.0748 & 0.0861 & 0.1134 & 0.0986 & 0.0921 & 0.1069 & 0.1204 & 0.1169 & \underline{0.1217} & \textbf{0.1319} \\
& NDCG@3 & 0.0528 & 0.0615 & 0.0769 & 0.1031 & 0.0852 & 0.0807 & 0.0961 & 0.1055 & 0.1033 & \underline{0.1088} & \textbf{0.1173} \\
& HR@5 & 0.0789 & 0.0865 & 0.0949 & 0.1252 & 0.1163 & 0.1048 & 0.1204 & 0.1406 & 0.1356 & \underline{0.1420} & \textbf{0.1496} \\
& NDCG@5 & 0.0595 & 0.0664 & 0.0805 & 0.1079 & 0.0960 & 0.0859 & 0.1017 & 0.1139 & 0.1110 & \underline{0.1172} & \textbf{0.1246} \\
& HR@10 & 0.1019 & 0.1093 & 0.1120 & 0.1400 & 0.1408 & 0.1237 & 0.1434 & \underline{0.1634} & 0.1587 & \underline{0.1634} & \textbf{0.1689} \\
& NDCG@10 & 0.0669 & 0.0737 & 0.0858 & 0.1126 & 0.1002 & 0.0920 & 0.1091 & 0.1213 & \underline{0.1255} & 0.1242 & \textbf{0.1308} \\
\bottomrule
\end{tabular}
\end{table*}

\stitle{Request-Specific Valid-Path Trie.}
{Let $\Omega$ denote the full creative inventory. To bound construction and filtering costs, we randomly sample a daily refreshed serving library $\Omega'\subset\Omega$ at the creative level and construct the global trie $\mathcal{T}$ over $\Omega'$. Each ad creative $i\in\Omega'$ is inserted along its complete CSID path $\Phi(i)=(s_i^1,\ldots,s_i^L)$. Shared SID prefixes reuse the same internal nodes, and each leaf $y$ stores its serving creative bucket $\mathcal{A}(y)\cap\Omega'$. For an incoming request $x$, targeting filters first remove creatives that violate advertiser-specified user conditions, such as gender, age, or region. Ad-level filters then remove creatives associated with unavailable ads, such as those with exhausted budgets or paused delivery. Finally, creative-level filters enforce constraints such as frequency capping, duplicate exposure, and content compliance. These filters produce the request-eligible inventory $\mathcal{I}(x)\subseteq\Omega'$.

We update each leaf bucket as $\mathcal{A}_x(y)=\mathcal{A}(y)\cap\mathcal{I}(x)$ and remove a leaf only when $\mathcal{A}_x(y)=\varnothing$. Internal nodes without remaining children are recursively pruned. The resulting subtree forms the request-specific trie $\mathcal{T}_x$, which retains exactly the CSID paths that can resolve to at least one deliverable creative.} Given a SID prefix $s_{<l}$, the valid next-token set is
\begin{equation}
\mathcal{V}(s_{<l}; \mathcal{T}_x) = \{a \in \mathcal{S}_l \mid (s_{<l},a) \in \mathcal{P}(\mathcal{T}_x)\},
\end{equation}
where $\mathcal{P}(\mathcal{T}_x)$ denotes the valid prefixes in the personalized trie. The trie defines a hard feasibility mask:
\begin{equation}
m_x(a\mid s_{<l}) =
\begin{cases}
0, & a \in \mathcal{V}(s_{<l}; \mathcal{T}_x), \\
-\infty, & \text{otherwise}.
\end{cases}
\end{equation}
It does not estimate commercial value. Its only role is to prevent request-invalid actions from consuming beam capacity.

\stitle{Value-Guided Personalized Beam Search.}
At decoding step $l$, UniVA computes the Generation-as-Ranking logits $\ell_{\mathrm{GAR}}^{(l)}$ from the current request and SID prefix. The feasibility mask then defines the serving distribution
\begin{equation}
\log\tilde{\pi}_x^{(l)}(a\mid s_{<l},h)
=\log\mathrm{Softmax}\left(
\ell_{\mathrm{GAR}}^{(l)}+m_x(\cdot\mid s_{<l})
\right)[a].
\end{equation}
Before each top-$B$ expansion, invalid actions receive zero probability and valid prefixes are ranked by cumulative log-probability:
\begin{equation}
\mathrm{Score}_x(s_{\leq l})
=\sum_{t=1}^{l}
\log\tilde{\pi}_x^{(t)}(s_t\mid s_{<t},h),
\qquad
s_{\leq l}\in\mathcal{P}(\mathcal{T}_x).
\end{equation}
Thus, the trie determines which actions are feasible, while the masked fused policy determines their order according to generation relevance and learned commercial return.

\stitle{Leaf-Level Ad Resolution.}
{After beam search selects a complete path $y$, the bid-priority resolver returns a concrete advertisement from its request-eligible bucket.} The trie guarantees that this bucket is nonempty. {Online} serving consequently performs one constrained autoregressive beam-search procedure that jointly filters invalid actions, generates SIDs, and ranks feasible trajectories. It does not run MCTS or invoke a separate learned value model. The personalized trie, Generation-as-Ranking decoder, and deterministic leaf resolver therefore avoid a global post-generation reranking stage.
}

\section{Experiments}
\subsection{Experimental Setup}

\stitle{Datasets.}
We evaluate UniVA on two public sequential recommendation benchmarks and a large-scale industrial advertising dataset. For public evaluation, we use the \textit{Industrial\_and\_Scientific} (Industrial) and \textit{Office\_Products} (Office) subsets of Amazon Reviews \cite{kong2025minionerec}. For industrial evaluation, following GPR \cite{zhang2025gpr}, we construct a large-scale Tencent advertising dataset containing mixed ad--organic traffic, session-level user behaviors, and multimodal item features. This dataset is used to evaluate the complete UniVA framework, including CSID and eCPM-aligned decoding. Detailed preprocessing procedures and descriptions are provided in Appendix~\ref{app:dataset_details}.

\stitle{Baselines.}
On the two public benchmarks, we compare UniVA with representative existing methods, including traditional sequential recommenders (GRU4Rec \cite{hidasi2016gru4rec}, Caser \cite{tang2018caser}, and SASRec \cite{kang2018sasrec}), generative recommenders (HSTU \cite{zhai2024actions}, TIGER \cite{rajput2023recommender}, and LCRec \cite{zheng2024lcrec}), and LLM-based recommenders (BIGRec \cite{bao2025bigrec}, D$^3$ \cite{bao2024d3}, S-DPO \cite{chen2024sdpo}, and MiniOneRec \cite{kong2025minionerec}). On the large-scale Tencent advertising dataset, we adopt GPR \cite{zhang2025gpr} as the primary system baseline and a vanilla decoder-only Transformer as the baseline SID decoder, upon which different UniVA components are progressively introduced. Additional implementation details are provided in Appendix~\ref{app:implementation_details}.

\stitle{Evaluation Metrics.} On the public Amazon benchmarks, we report HR@K and NDCG@K at $K\in\{3,5,10\}$. On the industrial advertising benchmark, we report HR@K for next-item interaction prediction. To evaluate commercial value on the GMV-weighted next-conversion set, we additionally report ValueHR@K, which measures the Top-$K$ coverage of conversion value, and wNDCG@K, which emphasizes the ranking quality of high-value conversions. For online evaluation, we report GMV and GMV(normal). Formal definitions and construction details of the value-oriented and online metrics are provided in Appendix~\ref{app:metric_details}.

\begin{table}[t]
\caption{Offline HR@100 ablation on the industrial advertising benchmark. {Rows are progressive configurations.} {Parameters and inference FLOPs cover only the SID decoder and exclude the encoder.} $\Delta$ denotes the relative improvement over the base.}
\label{tab:main}
\scalebox{0.97}{
\begin{tabular}{@{}llll@{}}
\toprule
\textbf{Model}                          & \textbf{Parameters} & {\textbf{FLOPs}} & \textbf{$\Delta$HR@100} \\ \midrule
\textbf{Base}                      &            &                                                                     &                         \\
\quad GPR+SID Decoder & 3M         & 4.1G                                                          & + 0.0\%                       \\ \midrule
\multicolumn{4}{@{}l}{\textbf{Tokenization and Decoder}} \\
\quad + CSID   & 3M         & 4.1G                                                         & +5.78\%                  \\
\quad {+ Deeper Decoder}  & 7M         & 7.1G                                                          & +6.10\%                  \\
\quad + MoR              & 5M         & 7.1G                                                          & +13.56\%                 \\
\quad + Sparse MoE       & 60M        & 8.5G                                                         & +18.40\%                 \\ \midrule
\textbf{RL Design}   &            &                                                                     &                         \\
{\quad + eCPM-aware RL} & {62M} & {8.5G} & {+32.01\%} \\
{\textbf{UniVA (Full)}}  & {{80M}} & {{23.2G}} & {\textbf{+37.04\%}} \\ \bottomrule
\end{tabular}}
\end{table}

\subsection{Overall Performance}
\stitle{Public Benchmark Results.} Table~\ref{tab:main_result} summarizes the comparison between UniVA and all baselines on the two public Amazon datasets. UniVA achieves the best performance across all reported metrics, with relative improvements ranging from 1.5\% to 8.4\% over the strongest baseline for each metric. On Amazon Industrial, the largest relative gains reach 2.9\% on HR and 3.0\% on NDCG. On Office, the corresponding gains reach 8.4\% on HR and 7.8\% on NDCG. {These consistent improvements show that UniVA's value-aware decoder generalizes to rating-based user-utility supervision, even though the public datasets use semantic SIDs without advertising-specific CSID construction.}

\stitle{Industrial Advertising Benchmark Results.} {GPR is an industrial baseline that incorporates commercial signals through value-aware training and decoding. Table~\ref{tab:main} evaluates UniVA's pipeline-wide alignment across SID representation, prefix-level learning, and request-constrained serving.} CSID improves HR@100 by 5.78\% without increasing decoder parameters or computation. This suggests that joint content--value grouping makes SID paths more commercially coherent and provides a clearer learning signal. Table~\ref{tab:main} also reveals a scaling trend for the SID decoder. As capacity increases, the HR@100 gain rises from 6.10\% for the deeper decoder to 13.56\% for MoR and further to 18.40\% for Sparse MoE. This suggests that SID decoding in industrial advertising benefits from stronger modeling capacity: MoR improves effective depth through recursive refinement, while Sparse MoE increases conditional capacity through expert specialization. Together, these results show that SID decoding scales with more expressive decoder backbones. {Detailed configurations are provided in Appendix~\ref{app:implementation_details}.}

{{The eCPM-aware RL configuration with Generation-as-Ranking decoding increases the relative HR@100 improvement from 18.40\% to 32.01\%.} This gain shows that terminal commercial feedback and action-value-guided token selection better preserve commercially promising SID trajectories than supervised next-SID learning alone. {Full UniVA increases the MoR recursion budget and scales the sparse MoE from 16 experts with Top-4 activation to 64 experts with Top-16 activation. This expanded configuration reaches a 37.04\% relative improvement, showing that value alignment benefits from increased effective depth and conditional capacity; detailed configurations are provided in Appendix~\ref{app:implementation_details}.} Overall, CSID, scalable decoding, and eCPM-aware Generation-as-Ranking provide complementary improvements on the industrial benchmark.}

\subsection{Value Alignment Performance}
\begin{figure}[tbp]
\centering
\includegraphics[width=0.95\columnwidth]{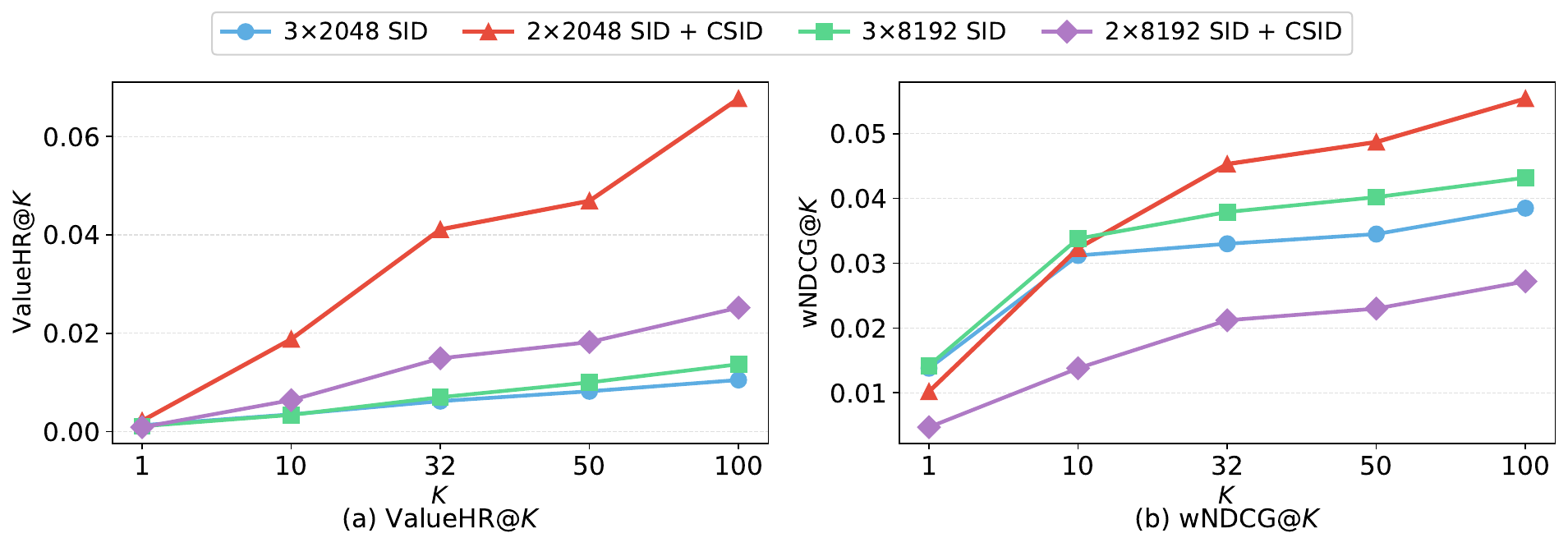}
\caption{Offline value analysis under four SID designs on the GMV-weighted next-conversion set.}
\label{fig:offline_value} 
\end{figure}

Figure~\ref{fig:offline_value} evaluates commercial-value capture under four SID designs within the same UniVA training and decoding framework. Across these configurations, eCPM-aware learning and Generation-as-Ranking provide a common mechanism for modeling and using commercial return, while their differences isolate the effect of SID construction. The $2\times2048$ SID + CSID configuration performs best at most cutoffs, achieving the highest ValueHR@10/32/50 and wNDCG@32/50; at $K=100$, it reaches 0.0677 ValueHR@100 and 0.0554 wNDCG@100. The semantic-only variants remain slightly stronger at the most restrictive positions, whereas CSID performs better at broader cutoffs by exposing commercial distinctions directly in the token space. Its advantage over $2\times8192$ SID + CSID further suggests that a moderate codebook avoids fragmented supervision and supports more stable value-aware decoding. Overall, the results validate both UniVA's value-aware learning framework and the complementary benefit of CSID.

\begin{figure}[tbp]
\centering
\includegraphics[width=0.95\columnwidth]{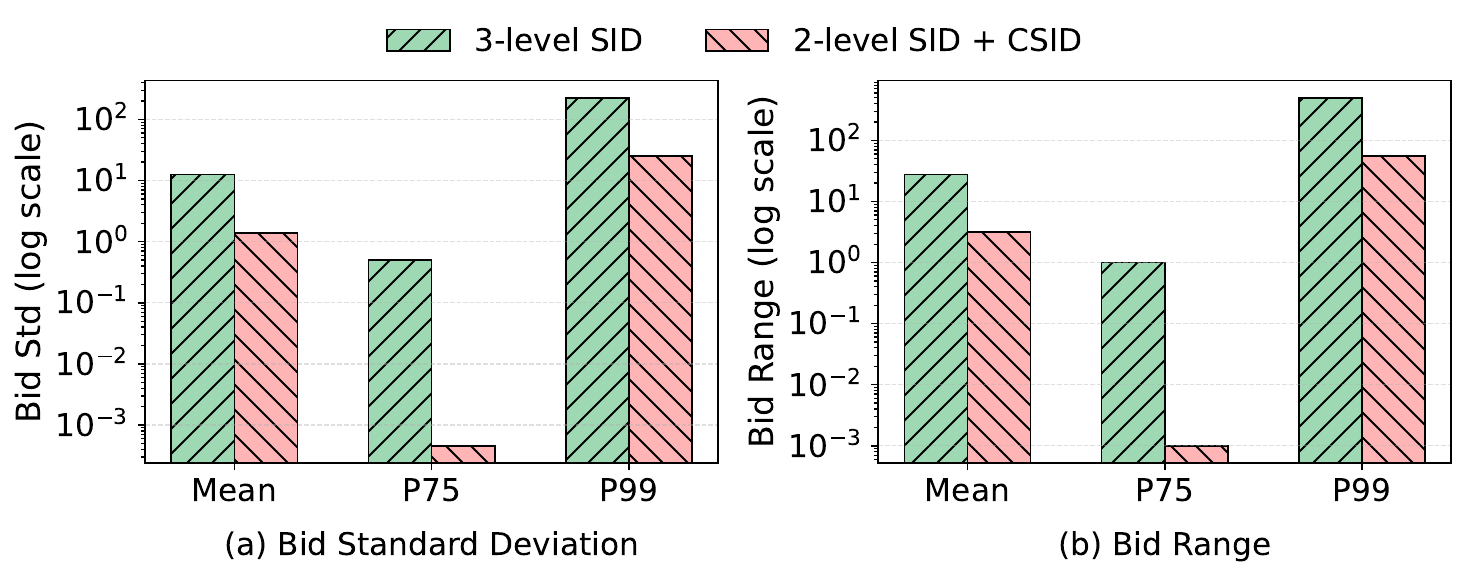}
\caption{Path-level bid-dispersion statistics for 3-level SID and 2-level SID + CSID. Each subfigure reports Mean, P75, and P99 over complete SID paths.}
\label{fig:csid_quality} 
\end{figure}

\subsection{More Insights}
\stitle{CSID Quality Analysis.} 
Figure~\ref{fig:csid_quality} shows that 2-level SID + CSID consistently reduces path-level bid standard deviation and range relative to 3-level SID across Mean, P75, and P99. Most statistics decrease by about one order of magnitude, especially in the middle and tail. This indicates that ads assigned to the same full SID path have more consistent commercial values, rather than mixing widely different bid levels under a shared semantic path. Thus, CSID produces commercially coherent paths and reduces high-variance groups, providing a cleaner basis for value-aware decoding. {Appendix~\ref{app:case_study} provides complementary case- and attribute-level analyses.}

\stitle{Codebook Size Analysis.}
Table~\ref{tab:codebook} compares HR@K across semantic codebook sizes and CSID settings. $2\times2048$ SID + CSID performs best at all cutoffs, with {relative improvements of 55.56\%, 41.67\%, 35.63\%, 32.09\%, and 30.03\% over $3\times2048$ SID, respectively.} Under the 8192 setting, however, $2\times8192$ SID + CSID remains below $3\times8192$ SID. The fixed 2048-way CSID vocabulary naturally matches the 2048 semantic setting, whereas replacing an 8192-way semantic level creates a larger mismatch. This mismatch fragments semantic supervision and weakens the benefit of fine-grained semantic partitioning. A larger semantic vocabulary is therefore not automatically beneficial when the commercial vocabulary remains fixed. Thus, CSID works best with a moderate codebook that balances semantic organization, commercial differentiation, and learning stability. Additional construction-strategy results appear in Appendix~\ref{app:csid_strategy}.

\begin{table}[t]

\caption{HR@K comparison under different SID codebook sizes. All values are reported in percentage points (\%).}
\label{tab:codebook}
\scalebox{0.8}{
\begin{tabular}{@{}llllll@{}}
\toprule
\textbf{SID Configuration} & \textbf{HR@1} & \textbf{HR@10} & \textbf{HR@32} & \textbf{HR@50} & \textbf{HR@100} \\ \midrule
$3\times2048$ SID        & 0.09 & 0.72  & 1.60  & 2.15  & 3.23   \\
$3\times8192$ SID        & 0.10 & 0.83  & 2.06  & 2.77  & 4.03   \\
$2\times2048$ SID + CSID & \textbf{0.14} & \textbf{1.02}  & \textbf{2.17}  & \textbf{2.84}  & \textbf{4.20}   \\
$2\times8192$ SID + CSID & 0.09 & 0.92  & 1.98  & 2.63  & 3.84   \\ \bottomrule
\end{tabular}}
\end{table}

\stitle{Trie-Constrained Serving Analysis.} {Using the same fused scorer and a beam width of 300, we compare global beam search followed by validity filtering with personalized trie-constrained beam search. The former returns only 48 valid SID paths, whereas the latter returns 300.} The average bid also increases from 1.29 to 11.44, nearly one order of magnitude. By pruning invalid branches before expansion, the trie prevents them from consuming limited beam capacity. This is particularly important under a fixed beam width, where each invalid expansion may displace a feasible candidate. The Generation-as-Ranking policy can therefore retain more deliverable, high-bid candidates. This confirms that constrained value-aware decoding improves retrieval validity and commercial quality.

\subsection{Online A/B Test}
{\stitle{Deployment Setting and Online Variants.} We conduct online A/B testing on Tencent WeChat Channels (Weixin Channels), a production advertising platform serving hundreds of millions of active users and tens of millions of dynamic ads. The experiment runs from March 7 to March 11, 2026, on 20\% of production traffic. Table~\ref{tab:online} compares four successive online versions against the same production baseline. V0 combines a semantic SID with GRPO. V1 retains this configuration and scales the SID decoder. V2 further replaces the semantic SID with CSID while keeping the scaled decoder and GRPO. Finally, V3 retains V2's CSID and scaled decoder, replaces GRPO with PPO-based action-value learning, and enables Generation-as-Ranking during online decoding.

\stitle{Online GMV Results.} V0 improves GMV and GMV(normal) by 0.22\% and 0.45\%, respectively, establishing the effectiveness of the GRPO-based generative baseline. Decoder scaling in V1 raises the two lifts to 0.36\% and 0.77\%, showing that increased model capacity improves production SID generation. With the same scaled decoder and GRPO objective, V2 introduces CSID and reaches gains of 1.03\% and 1.17\%. The substantial increase over V1 demonstrates that exposing commercial distinctions in the token space benefits online monetization. Full UniVA achieves the strongest performance, with a 1.50\% GMV lift and a 1.42\% GMV(normal) lift. Compared with V2, this final gain reflects the combined effect of PPO-based prefix-level action-value learning and its reuse through Generation-as-Ranking, confirming that value-aware token selection translates into additional business value without a separate online reranker.}

\begin{table}[t]
\caption{Online A/B test results on Tencent WeChat Channels advertising traffic from March 7 to March 11, 2026, using 20\% of production traffic. All lifts denote relative improvement against the production baseline.}
\label{tab:online}
{
\begin{tabular}{@{}lll@{}}
\toprule
\textbf{Online Version}                & \textbf{GMV Lift}    & \textbf{GMV(normal) Lift} \\ \midrule
{V0: SID + GRPO} & +0.22\% & +0.45\% \\
{V1: V0 + Scaling} & +0.36\% & +0.77\% \\
{V2: V1 + CSID} & +1.03\% & +1.17\% \\
{V3: UniVA} & \textbf{+1.50\%} & \textbf{+1.42\%} \\ \bottomrule
\end{tabular}}
\end{table}

\section{Conclusion}
This paper identifies value inconsistency across SID representation, autoregressive decoding, and online serving, and proposes UniVA to align commercial value throughout the generative advertising pipeline. CSID Tokenization constructs a value-discriminative token space, while the Generation-as-Ranking SID Decoder learns generation likelihoods and prefix-level action values. During serving, Value-Aware Constrained Serving fuses these signals over request-valid paths. UniVA improves offline HR@100 by 37.04\% and online GMV by 1.50\%, validating pipeline-wide value alignment.

\bibliographystyle{ACM-Reference-Format}
\bibliography{sample-base}

\appendix
\newpage
\section{Related Work}
\label{app:related_work}

\stitle{Generative Recommendation.}
GR reformulates recommendation as autoregressive generation over discrete item identifiers, thereby replacing hand-crafted multi-stage ranking stacks with a unified next-token prediction paradigm \cite{rajput2023recommender,zhai2024actions,zhou2025onerec,yi2025recgpt,han2025mtgr}. Recent advances show that this line is evolving quickly from proof-of-concept models to industrial-scale systems, with active exploration of one-model training, large-catalog generation, decoding acceleration, and scalable serving architectures \cite{wang2025nezha,zhou2025onerec2,huang2025genrank}. The same trend has reached advertising: GPR establishes a generative pre-trained one-model architecture with business-aware training and decoding, while EGA-v2 and GR4AD further explore end-to-end generative advertising systems \cite{zhang2025gpr,zheng2025ega,xue2026GR4AD}. These efforts demonstrate both the feasibility of industrial GR and the importance of commercial signals. UniVA instead focuses on aligning those signals consistently across SID representation, prefix-level decoding, and request-constrained serving.

\stitle{Semantic ID.}
Semantic ID is a key enabler of large-scale generative recommendation because it converts items in massive catalogs into compact discrete token sequences. Early work learns vector-quantized item identifiers from semantic embeddings \cite{hou2023learning}, while later studies improve the SID pipeline through learnable tokenization, large-catalog training, and parallel generation of long semantic-ID sequences \cite{petrov2024recjpq,wang2024letteer,hou2025generating}. More recent work further moves toward end-to-end SID generation in industrial advertising environments \cite{jiang2026unisid}. Although these studies substantially improve semantic fidelity, token efficiency, and generation scalability, they are still primarily semantics-driven: the SID space is designed to preserve item similarity or facilitate generation, rather than to explicitly expose monetization differences.

\stitle{Value Modeling in Recommendation.}
Value-aware recommendation has long recognized that commercial systems should optimize monetization-related utility in addition to user relevance, leading to research on profit-aware learning, multi-objective recommendation, and value alignment in industrial scenarios \cite{pei2019value,luo2025qarm}. In advertising, recent generative systems have also started to incorporate auction signals, eCPM-oriented supervision, and business-aware objectives into model training and serving \cite{zhang2025gpr,zheng2025ega,zhao2025llm,xue2026GR4AD,xiong2026llatte}. However, these signals are generally introduced through stage-specific mechanisms, leaving value alignment incomplete across SID representation, prefix-level decoding, and request-constrained serving. UniVA differs by treating value as a pipeline-wide modeling principle.

\section{Optimization Details}
\label{app:optimization_details}

{\stitle{Why PPO instead of GRPO.}
GRPO estimates group-relative advantages from sampled rewards without learning an explicit critic \cite{shao2024deepseekmath}. Although this design is effective for complete-trajectory preference optimization, standard GRPO does not explicitly produce a request-conditioned evaluator for intermediate prefix--action pairs. UniVA requires such estimates to guide offline MCTS exploration and to rank candidate token expansions during online Generation-as-Ranking. We therefore adopt PPO \cite{schulman2017proximal} with action-value regression, while the short horizon of advertising SIDs makes explicit prefix-level value learning computationally practical.}

\stitle{Terminal Reward Construction.}
For each complete feasible trajectory $y^{(k)}$, the bid-priority rule
in Eq.~\ref{eq:leaf_resolution} resolves its leaf bucket to a concrete
advertisement. The offline simulator then returns
\begin{equation}
R^{(k)}_{\mathrm{eCPM}}
=g_{\mathrm{eCPM}}\big(h,\rho_x(y^{(k)})\big).
\end{equation}
To preserve reward fidelity, $g_{\mathrm{eCPM}}$ uses the predicted click-through rate (pCTR) and predicted conversion rate (pCVR) components copied from the production ranking stack. These frozen components score the resolved advertisement under the current request, aligning the offline reward with the serving-time monetization objective rather than a handcrafted proxy.
For each request $x$, let
$\{R_{\mathrm{eCPM}}^{(1)},\ldots,R_{\mathrm{eCPM}}^{(K_{\mathrm{traj}})}\}$
be the simulator rewards of the collected trajectories. We normalize
these rewards within the request:
\begin{equation}
\mu_R(x)=\frac{1}{K_{\mathrm{traj}}}\sum_{k=1}^{K_{\mathrm{traj}}}R_{\mathrm{eCPM}}^{(k)}, \qquad
\sigma_R(x)=
\sqrt{\frac{1}{K_{\mathrm{traj}}}\sum_{k=1}^{K_{\mathrm{traj}}}
\left(R_{\mathrm{eCPM}}^{(k)}-\mu_R(x)\right)^2},
\end{equation}
\begin{equation}
\bar{R}^{(k)}
=\frac{R_{\mathrm{eCPM}}^{(k)}-\mu_R(x)}
{\sigma_R(x)+\epsilon_r},
\end{equation}
where $\epsilon_r>0$ ensures numerical stability. The normalized
reward is assigned only when a complete SID path resolves to an
advertisement:
\begin{equation}
r_l^{(k)}=0 \quad (l<L), \qquad
r_L^{(k)}=\bar{R}^{(k)}.
\label{eq:sparse_terminal_reward}
\end{equation}
This construction reduces reward-scale variation across traffic
contexts and makes the update depend on relative commercial value
among trajectories for the same request.

\stitle{Action-Value-Guided MCTS.}
For a search node $n$ corresponding to prefix $s_n$, let
$\mathcal{V}(n)$ be its valid next-token set, and let $N(n)$ and
$N(n,a)$ denote the node and edge visit counts. Let $\bar{Q}(n,a)$
denote the backed-up return estimate for edge $(n,a)$ and
$c_{\mathrm{puct}}>0$ the exploration coefficient. UniVA uses the frozen
generation policy as the search prior and selects an action according
to the {Predictor + Upper Confidence bounds applied to Trees (PUCT)} rule \cite{silver2017mastering}:
\begin{equation}
a^\star=\arg\max_{a\in\mathcal{V}(n)}
\left[
\bar{Q}(n,a)
+c_{\mathrm{puct}}\,
\pi_{\mathrm{old}}(a\mid x,s_n)
\frac{\sqrt{N(n)}}{1+N(n,a)}
\right].
\end{equation}
The frozen action-value head supplies the estimate for an unvisited
edge. Once the edge has been evaluated by one or more completed
simulator trajectories, their normalized terminal rewards are backed
up as an empirical mean:
\begin{equation}
\bar{Q}(n,a)=
\begin{cases}
q_{\mathrm{old}}(s_n,a;x), & N(n,a)=0,\\[2mm]
\displaystyle
\frac{1}{N(n,a)}
\sum_{\tau\in\mathcal{R}(n,a)}\bar{R}^{(\tau)},
& N(n,a)>0,
\end{cases}
\end{equation}
where $\mathcal{R}(n,a)$ is the set of indices of completed trajectories
whose search paths traverse edge $(n,a)$, and $\tau$ indexes a trajectory
in this set. For a visited edge, $N(n,a)=|\mathcal{R}(n,a)|$. This design uses the same
action-value estimates later retained by Generation-as-Ranking,
rather than introducing a separate MCTS evaluator.

\stitle{Advantage Estimation and PPO Objective.}
For a prefix $s_{<l}$, let
$\mathcal{V}_l(x,s_{<l})\subseteq\mathcal{S}_l$ denote its valid
next-token set. We renormalize a policy over this support as
\begin{equation}
\pi_{\vartheta}^{\mathcal{V}}(a\mid x,s_{<l})
=
\frac{\mathbb{I}\!\left[a\in\mathcal{V}_l(x,s_{<l})\right]
\pi_{\vartheta}(a\mid x,s_{<l})}
{\sum_{a'\in\mathcal{V}_l(x,s_{<l})}
\pi_{\vartheta}(a'\mid x,s_{<l})},
\qquad
\vartheta\in\{\theta,\mathrm{old}\}.
\end{equation}
The frozen action-value head induces the behavior-policy state
baseline
\begin{equation}
V_l^{\mathrm{old}}
=
\sum_{a\in\mathcal{V}_l(x,s_{<l})}
\pi_{\mathrm{old}}^{\mathcal{V}}(a\mid x,s_{<l})
q_{\mathrm{old}}(s_{<l},a;x),
\qquad
V_{L+1}^{\mathrm{old}}=0.
\end{equation}
Given the sparse rewards in Eq.~\ref{eq:sparse_terminal_reward}, generalized advantage
estimation computes
\begin{equation}
\delta_l
=r_l+\gamma V_{l+1}^{\mathrm{old}}-V_l^{\mathrm{old}},
\qquad
\widehat{A}_l
=\operatorname{sg}\!\left(
\sum_{j=0}^{L-l}(\gamma\lambda)^j\delta_{l+j}
\right),
\end{equation}
\begin{equation}
\widehat{G}_l
=\operatorname{sg}\!\left(
\widehat{A}_l+V_l^{\mathrm{old}}
\right),
\end{equation}
where $\operatorname{sg}(\cdot)$ stops gradients through the targets.
The valid-support importance ratio is
\begin{equation}
\rho_l(\theta)
=
\frac{\pi_\theta^{\mathcal{V}}(a_l\mid x,s_{<l})}
{\pi_{\mathrm{old}}^{\mathcal{V}}(a_l\mid x,s_{<l})}.
\end{equation}
The clipped policy loss is
\begin{equation}
\mathcal{L}_{\mathrm{PPO}}
=-\mathbb{E}\left[
\sum_{l=1}^{L}
\min\!\left(
\rho_l(\theta)\widehat{A}_l,\,
\operatorname{clip}\!\left(
\rho_l(\theta),1-\epsilon,1+\epsilon
\right)\widehat{A}_l
\right)
\right].
\end{equation}
The action-value term in the main-text RL objective regresses the
selected token value $q_\phi(s_{<l},a_l;x)$ toward
$\widehat{G}_l$. Consequently, PPO updates the generation policy,
while return regression trains the vocabulary-level action values
reused by MCTS and online Generation-as-Ranking.

\section{Additional CSID Analysis}
\label{app:csid_strategy}

\begin{figure}[tbp]
\centering
\includegraphics[width=0.85\columnwidth]{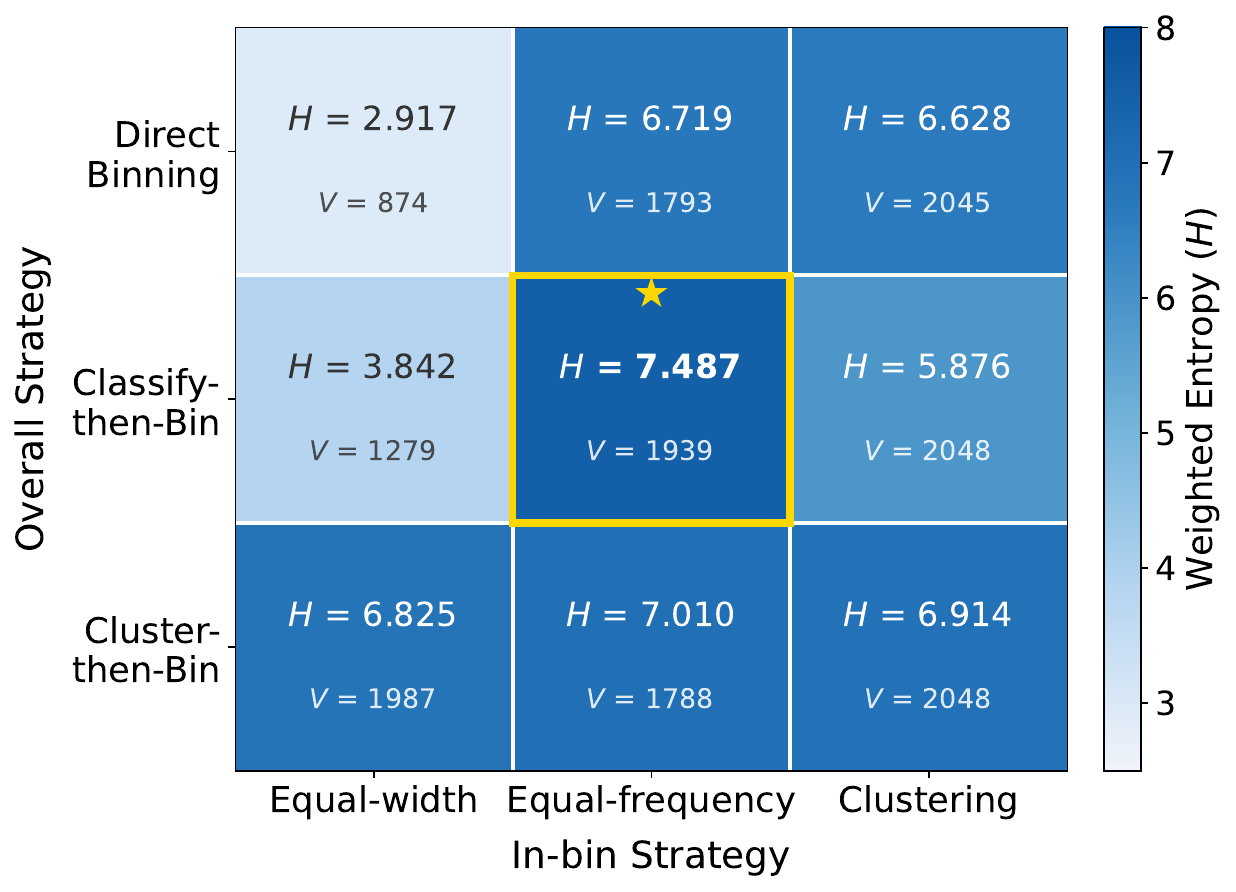}
\caption{CSID strategy comparison across three overall strategies and three in-bin strategies. Each cell reports commercial-token entropy $H_{\mathrm{tok}}$ and vocabulary size $V$.}
\label{fig:csid_strategy}
\end{figure}

\stitle{Construction Strategy Analysis.}
Figure~\ref{fig:csid_strategy} compares different CSID construction strategies. Classify-then-Bin combined with equal-frequency binning achieves the highest weighted entropy while keeping the vocabulary size closest to the target budget of 2048, with $H_{\mathrm{tok}} = 7.487$ and $V = 1939$. Direct Binning ignores structured commercial attributes and therefore mixes heterogeneous ads before bid discretization, leading to coarse and less balanced partitions. Cluster-then-Bin improves bid-distribution grouping, but the resulting clusters are less stable and often trade vocabulary efficiency for limited entropy gain. Among in-bin strategies, equal-width binning is sensitive to long-tail bid distributions, while clustering tends to consume more vocabulary without consistent benefit. Overall, the results support Classify-then-Bin with equal-frequency binning as the most balanced strategy for CSID construction.

{
\section{Case Study}
\label{app:case_study}

Figure~\ref{fig:csid_case_study} compares a conventional 3-level semantic SID with UniVA's 2-level SID + CSID on ten game advertisements. The semantic SID maps all ads to $[0,1071,307]$ despite four distinct bid values, yielding a within-path bid range of $0.09$. CSID retains the prefix $[0,1071]$ but separates the bid groups into four commercial leaves, each with zero within-path bid range. Thus, CSID improves commercial discriminability without increasing sequence length or disrupting the shared semantic hierarchy.
 
\begin{figure}[t]

\centering
\includegraphics[width=0.92\linewidth]{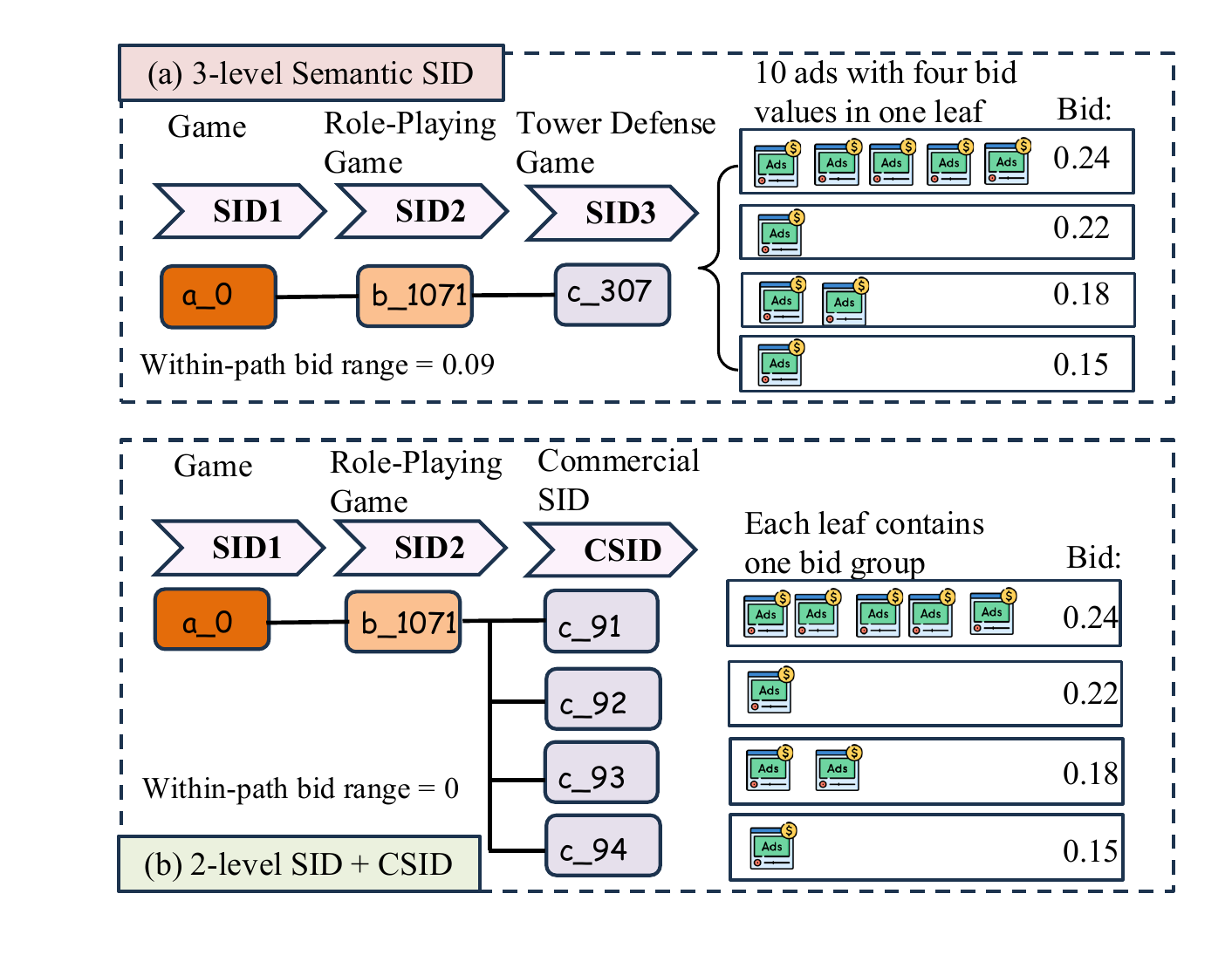}
\caption{Case study comparing a 3-level semantic SID with 2-level SID + CSID at the same sequence length. Ten game advertisements share the semantic path $[0,1071,307]$ but have four bid values. CSID preserves the semantic prefix $[0,1071]$ and assigns commercial leaf tokens $91$--$94$, reducing the within-path bid range from $0.09$ to $0$. Commercial token IDs are categorical.}
\Description{A comparison of a three-level semantic SID and a two-level semantic SID plus a Commercial SID. The semantic SID places ten advertisements with four bid values in one leaf, whereas CSID separates them into four commercial leaves with one bid group per leaf.}
\label{fig:csid_case_study}
\end{figure}

\stitle{Commercial Attribute Homogeneity.}
We further evaluate 691,317 advertisements randomly sampled from the industrial dataset. Both representations share the same two-level semantic prefix and differ only in the final token: the third semantic code for the original SID versus the commercial token for CSID. Figure~\ref{fig:commercial_attribute_homogeneity}(a) shows that CSID raises within-prefix conditional $R^2$ from 57.6\% to 99.2\% for effective bid and from 51.2\% to 89.9\% for expected ROI. Figure~\ref{fig:commercial_attribute_homogeneity}(b) further shows that CSID preserves the high CRM-industry and product-type leaf purity while improving optimization-goal, ROI-goal, and smart-bid-type purity. CSID therefore produces more commercially homogeneous leaves without sacrificing coarse semantic organization.

\begin{figure}[t]

\centering
\includegraphics[width=0.95\linewidth]{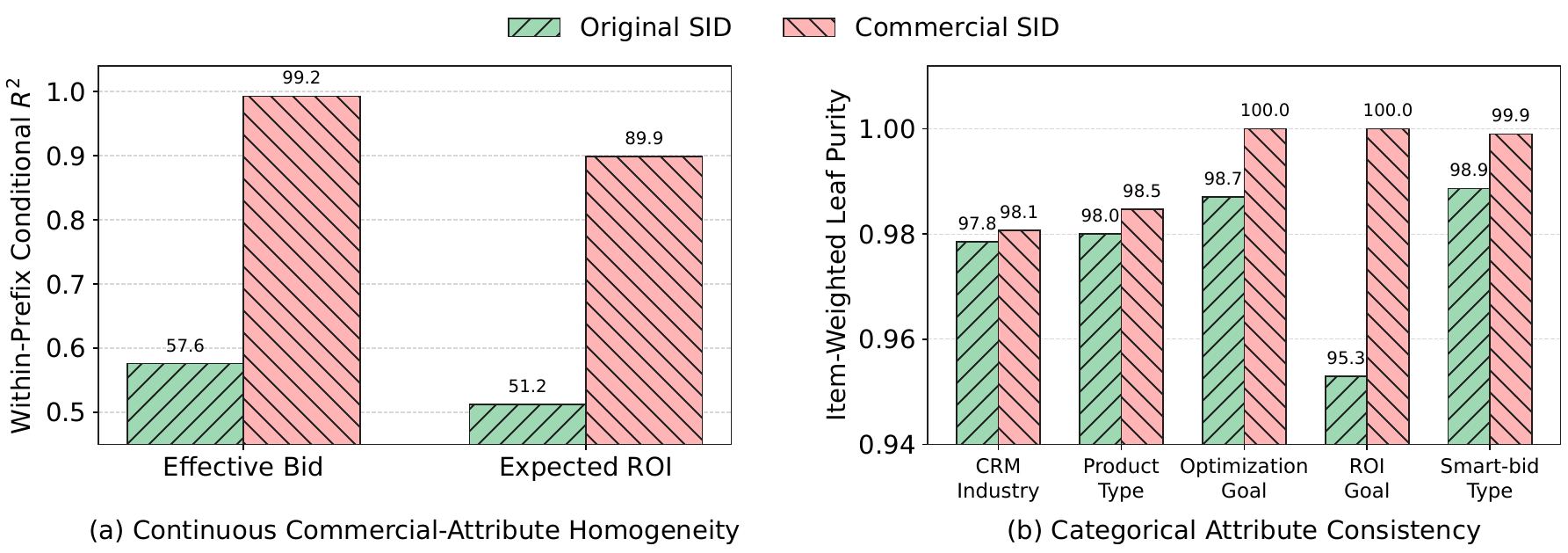}
\caption{Commercial-attribute homogeneity of the original 3-level semantic SID and 2-level SID + CSID, computed on 691,317 advertisement samples randomly drawn from the industrial dataset. (a) Within-prefix conditional $R^2$ measures the continuous-attribute variance explained by the final SID token after controlling for the shared two-level semantic prefix. (b) Item-weighted leaf purity measures categorical agreement within complete SID paths. Higher values indicate more homogeneous leaves.}
\Description{Two grouped bar charts compare the original semantic SID with Commercial SID. Commercial SID substantially improves conditional R-squared for effective bid and expected ROI, and achieves equal or higher leaf purity across CRM industry, product type, optimization goal, ROI goal, and smart-bid type.}
\label{fig:commercial_attribute_homogeneity}
\end{figure}
}

\section{Discussion and Limitations}
\label{app:discussion_limitations}

\stitle{Unified Action-Value Head.}
UniVA uses one vocabulary-level action-value head across policy learning, trajectory collection, and online serving. {During eCPM-aware policy optimization, selected action values are trained against return targets, while their behavior-policy expectation forms the state baseline used by PPO.} During offline MCTS collection, the same action values guide exploration of intermediate SID branches. During online serving, the same trained head is evaluated on the current request and fused with the corresponding generation logits. These stages share the action-value parameters and scoring function rather than stored logits or request data. MCTS remains offline, while online serving uses only the generation head, the reused action-value head, and personalized beam search.

\stitle{Limitations.}
First, UniVA relies on structured commercial attributes and simulator-provided eCPM feedback. Noisy reward models or distribution shifts in auctions may therefore weaken value alignment, requiring periodic recalibration and retraining. Public recommendation datasets also lack direct commercial labels, so their rating-based value proxy cannot fully reproduce the industrial setting. Second, CSIDs are constructed offline, whereas bids, inventory, and advertiser strategies evolve over time. Updating commercial partitions without destabilizing learned identifiers remains an important direction for incremental tokenization. Third, the current objective focuses on relevance and short-term commercial return. Broader goals such as long-term user satisfaction, advertiser fairness, and marketplace diversity are not explicitly optimized. The online evaluation is also conducted on a single production platform, and validation across other advertising markets and auction mechanisms is left for future work.

\section{Additional Experimental Details}
\label{app:experimental_details}

\subsection{Implementation Details}
\label{app:implementation_details}

\stitle{CSID Configuration.}
UniVA uses a three-level SID structure with a codebook size of 2048. On the industrial advertising dataset, we retain optimization-goal values covering 99\% of the samples and cluster the remaining values according to their bid distributions, resulting in 25 categories. For ROI, values covering 99\% of the samples are retained, while the remaining values are assigned to a fallback category, resulting in 8 categories. For industry, UniVA retains the top 9 first-level categories, which cover 75\% of the samples, and merges the remaining categories into a fallback group, resulting in 10 categories. {Composition keys with fewer than $m_{\min}=3$ samples are merged into the fallback key. The commercial level uses adaptive bid binning with $b_{\min}=3$ and $b_{\max}=25$ bins per retained key.} The final configuration is selected by grid search to maximize commercial-token entropy $H_{\mathrm{tok}}$ under a vocabulary budget of 2048.

{\stitle{Decoder Ablation Configurations.}
The base SID decoder contains two Transformer layers with an embedding dimension of 256, and adding CSID changes only the tokenization. The Deeper Decoder expands the backbone to four independent layers. MoR retains four effective layers through recursive parameter sharing with a recursion budget of $R=2$. Sparse MoE further replaces the dense feed-forward transformation with 16 routed experts, activates the top 4 experts per token, and uses an expert hidden dimension of 128. These variants correspond to the progressive configurations in Table~\ref{tab:main}.}

\stitle{Model and Training Configuration.}
{The 62M-parameter eCPM-aware RL configuration uses an embedding dimension of 256 and four effective decoder layers with an MoR recursion budget of $R=2$. Its sparse MoE contains 16 routed experts and activates the top 4 experts for each token. The hidden dimension of each expert is 128.} In training, UniVA optimizes the model with Adam using a learning rate of 0.001, and the batch size is 16. The input sequence length is 2048. {On the public Amazon benchmarks, which do not expose advertising-specific attributes, we retain semantic SIDs and do not construct the commercial SID level. Interaction-level ratings are used only as a user-satisfaction proxy for value supervision.}

{\stitle{Full UniVA Configuration.}
Full UniVA retains the embedding dimension of 256 and the expert hidden dimension of 128. It increases the effective decoder depth from four to six by raising the MoR recursion budget from $R=2$ to $R=4$. Because the middle transformation is shared across recursions, this depth expansion primarily increases computation rather than allocating an independent parameter set at every effective layer. In parallel, the sparse MoE is scaled from 16 to 64 routed experts, and the number of activated experts per token increases from 4 to 16. The resulting decoder contains 80M parameters and requires 23.2G inference FLOPs, compared with 62M parameters and 8.5G FLOPs for the eCPM-aware RL configuration.}

{\stitle{Serving Candidate Library.}
Let $\Omega$ denote the full creative inventory. At the creative level, we randomly sample a serving subset $\Omega'\subset\Omega$ containing approximately five million creatives and refresh it daily. The global trie $\mathcal{T}$ is rebuilt over the complete CSID paths of creatives in $\Omega'$. For each request $x$, the targeting, ad-level, and creative-level filters described in Section~\ref{sec:value_aware_serving} produce an eligible inventory $\mathcal{I}(x)\subseteq\Omega'$, from which the request-specific trie $\mathcal{T}_x$ is obtained.}

\subsection{Dataset Details}
\label{app:dataset_details}

\stitle{Public Datasets.}
We use the \textit{Industrial\_and\_Scientific} (Industrial) and
\textit{Office\_Products} (Office) subsets of Amazon Reviews.
Following the preprocessing protocol of MiniOneRec
\cite{kong2025minionerec}, we remove users and items with fewer than
five interactions and truncate each user interaction sequence to at
most ten items. The interactions of each user are sorted
chronologically and divided into training, validation, and test sets
with a ratio of 8:1:1. After preprocessing, the Industrial dataset
contains 3,685 items and 36,259/4,532/4,533
training/validation/test instances, respectively. The Office dataset
contains 3,459 items and 38,924/4,866/4,866
training/validation/test instances, respectively. Amazon Reviews does not provide advertising-specific attributes such as bid, optimization goal, ROI, and industry, nor an online eCPM reward model. {We therefore use only semantic SIDs on these datasets and do not construct CSIDs. The observed interaction-level rating is used exclusively as a user-satisfaction value signal for training the value-aware decoder; it does not participate in SID construction.} This signal represents user utility rather than commercial return. The public experiments evaluate the transferability of UniVA beyond advertising, whereas the industrial benchmark evaluates its full commercial value-alignment pipeline.

\subsection{Evaluation Metric Details}
\label{app:metric_details}

For the GMV-weighted next-conversion set, let $T$ be the number of evaluation requests, $i_t$ the ground-truth converted item for request $t$, and $R_t^K$ the model's Top-$K$ candidate set. ValueHR@K weights each successful retrieval by the GMV of the converted item:
\begin{equation}
\mathrm{ValueHR@K} = \frac{\sum_{t=1}^{T} \mathrm{gmv}_{i_t}\cdot\mathbb{I}(i_t\in R_t^K)}{\sum_{t=1}^{T}\mathrm{gmv}_{i_t}}.
\end{equation}
We further use wNDCG@K to emphasize the ranking positions of high-value conversions, with logarithmic GMV weights:
\begin{equation}
\mathrm{wNDCG@K} = \frac{\sum_{t=1}^{T}w_t\cdot\mathrm{NDCG}_t@K}{\sum_{t=1}^{T}w_t}, \qquad w_t=\log_{10}(1+\mathrm{gmv}_{i_t}).
\end{equation}
{
\stitle{Online GMV Metrics.}
Gross merchandise value (GMV) is the total value of completed transactions attributed to ads served during the online experiment. GMV(normal) uses the same attribution protocol but excludes return-on-investment (ROI) ads, reducing sensitivity to traffic-mix changes from campaigns that explicitly optimize ROI targets. Both metrics are reported as relative lifts over the contemporaneous production control:
\begin{equation}
\mathrm{Lift}(M)=\frac{M_{\mathrm{test}}-M_{\mathrm{control}}}{M_{\mathrm{control}}},
\qquad M\in\{\mathrm{GMV},\mathrm{GMV(normal)}\},
\end{equation}
where each metric is normalized by its allocated experiment traffic before computing the lift. GMV captures total commercial impact, whereas GMV(normal) provides a complementary view less sensitive to ROI-campaign composition.
}

\stitle{Industrial Advertising Dataset.}
Following GPR \cite{zhang2025gpr}, we construct the industrial offline
dataset from a large-scale Tencent advertising corpus collected from
mixed-traffic recommendation scenarios. The corpus contains both
advertisements and organic content, including short videos, social
feeds, and news, thereby reflecting the serving environment in which
advertising and organic content are jointly presented. Each sample
contains session-level user behaviors and item-level multimodal
features, including textual metadata and visual representations
extracted from covers or sampled video frames. For advertisements, the
dataset additionally provides commercial attributes, including
optimization goal, ROI target, industry, and bid, which are used to
construct CSIDs and provide value-aware supervision. We
remove near-duplicate samples, rebalance category distributions to
reduce sampling bias, and divide the resulting corpus into 80\%
training data and 20\% test data. Unlike the public datasets, this
industrial dataset supports evaluation of the complete UniVA
framework, including CSID tokenization, eCPM-aware
reinforcement learning, and value-guided decoding.
\end{document}